\documentclass[intlimits,twoside,a4paper]{article}

\usepackage[cp1251]{inputenc}
\usepackage{xcolor}
\usepackage[eqsecnum]{cmpj3}

\usepackage{bm}

\newcommand{\el}{\ell}

\newcommand{\inul}{\ri\nu_{\el}}

\issue{2020}{23}{4}{43703}
\doinumber{10.5488/CMP.23.43703}
\title[A pedagogical derivation of dynamical susceptibilities]%
{A pedagogical derivation of dynamical susceptibilities}
\author[R.D. Nesselrodt,  J.K. Freericks]{R.D. Nesselrodt,  J.K. Freericks}
\address{
Department of Physics, Georgetown University, 37th and O Sts. NW, Washington, DC 20057, U.~S.~A. 
}
\date{Received June 16, 2020, in final form July 13, 2020}
\begin{document}

\maketitle

\begin{abstract}
Dynamical two-particle susceptibilites are important for a wide range of different experiments in condensed-matter physics and beyond. Nevertheless, most textbooks avoid describing how to derive such response functions, perhaps because they are viewed as too complex. In the literature, most derivations work with generalized susceptibilities, which are more general, but require an even higher layer of complexity. In this work, we show a more direct derivation in the context of model Hamiltonians which can be mapped directly onto an impurity model. We also present an alternative derivation for the irreducible vertex in the context of the Falicov-Kimball model.

\keywords dynamical susceptibility, Keldysh formalism, linear response 
%
\end{abstract}

\section{Introduction}

One of the authors (J.~K.~F.), met Andrij Shvaika in 2000 at a conference in Lviv. He actually met him virtually before the meeting, serving as a referee for a conference proceeding Andrij had submitted earlier in the year. The initial meeting blossomed into a two-decades-long collaboration involving numerous funded grants, over two dozen joint publications, and a lifelong friendship. Time and again during our collaborative work, we would each approach problems from different perspectives. The western approach emphasized equations of motion and functional methods, while the eastern approach emphasized diagrammatic expansions based on Larkin irreducible parts of Green's functions. This worked well, because we usually were able to derive results from two different perspectives --- when they agreed, we had good confidence that they were correct.

About fifteen years ago, J.~K.~F.~developed some notes on how to derive dynamical susceptibilities using a direct nonequilibrium perturbation theory approach, but they were never published anywhere. It seemed appropriate to dust them off now and take the opportunity to write an article that does this here and that emphasizes the pedagogical aspect of the work. We then lost the notes during the COVID-19 pandemic and ended up reconstructing them in more detail; the result of these efforts is presented here. We hope it will help others who are looking for simpler ways to understand and work with dynamical susceptibilities.

A dynamical susceptibility satisfies the so-called Bethe-Salpeter equation. This is the two-particle equivalent of the Dyson equation for single-particle Green's functions.
The Bethe-Salpeter equation was first derived by Bethe and Salpeter in 1951 \cite{Bethe}.  This equation typically has a matrix character to it, but we will be deriving a vector character form of the equation here. Our approach is to begin with a static Hamiltonian and add a time-dependent disturbance to it. We then seek to determine the linear-response under such time-dependent driving.  We focus on a single-site lattice impurity;  one can then map the impurity to a lattice problem via the dynamical mean-field theory approach, but we do not provide all of those details. We also concentrate on treating spinless fermions, since we are working with the charge susceptibility and Andrij Shvaika has worked on many different spinless fermion systems in his career. Of course, our Bethe-Salpeter equation derivation can be easily generalized to more complex cases.

In many-body physics, we need to evaluate expectation values of different types of operator expressions. Within dynamical mean-field theory, we do so in the presence of a dynamical mean-field (denoted by $\lambda$), which represents how electrons hop onto and off  a given lattice site as a function of time. A themodynamic expectation value of a time-independent operator $\hat{A}$ is then given by
\begin{equation}
    \langle\hat{A}\rangle = \frac{1}{\mathcal{Z}}\text{Tr}\{\re^{-\beta \hat{\mathcal{H}}}\hat{S}(\lambda)\hat{A}\},
\end{equation}     
where $\mathcal{Z}=\text{Tr}\{\re^{-\beta \hat{\mathcal{H}}}\hat{S}(\lambda)\}$ is the partition function, $\hat{S}(\lambda)$ is a time-dependent evolution operator (describing the dynamical mean-field), and Boltzmann's constant is set equal to one. We will be examining the class of Hamiltonians which can be mapped to the impurity Hamiltonian, $\hat{\mathcal{H}}=-\mu\hat{n}$ in the presence of the dynamical mean-field $\lambda(\tau,\tau')$. This constitutes the set of Hamiltonians that can be treated via dynamical mean-field theory.

Here, $\hat{n}=\hat{c}^{\dagger}\hat{c}$, and we work in the interaction representation where 
\begin{equation}
\label{timev}
    S(\lambda)=\mathcal{T}_{\tau}\exp\bigg[-\int_0^{\beta}\,\rd\tau\ \int_0^{\beta}\,\rd\tau' \lambda(\tau,\tau')\hat{c}^{\dagger}(\tau)\hat{c}(\tau')\bigg]
\end{equation}
with $\mathcal{T}_{\tau}$ the (imaginary) time-ordering operator and $\beta$ the inverse temperature; all operators evolve in time according to the unperturbed Hamiltonian $\mathcal{H}$. The fermionic operators are ${\hat c}$ and ${\hat c}^\dagger$, which satisfy the canonical anticommutation relation $\{\hat c,\hat c^\dagger\}=1$. We have $\hat c(\tau)=\re^{\tau\hat{\mathcal H}}\hat c\re^{-\tau\hat{\mathcal H}}$ and $\hat c^\dagger(\tau)=\re^{\tau\hat{\mathcal H}}\hat c^\dagger \re^{-\tau\hat{\mathcal H}}$. Note that $\hat c^\dagger(\tau)\ne [\hat c(\tau)]^\dagger$ and $\hat n(\tau)=\hat n$.

Next, we define the Green's function in imaginary time as 
\begin{equation}
    G(\tau,\tau')=-\langle \mathcal{T}_{\tau}\hat{c}(\tau)\hat{c}^{\dagger}(\tau')\rangle
    =-\frac{1}{\mathcal Z}\text{Tr}\left\{\re^{-\beta \hat{\mathcal H}}\mathcal{T}_{\tau}\hat{S}(\lambda)\hat{c}(\tau)\hat{c}^{\dagger}(\tau')\right\}.
\end{equation}
Note that we do not assume, nor can one show, that the Green's function is time-translation invariant when the dynamical mean-field is not time-translation invariant (which occurs when we add a time-dependent field to the Hamiltonian). Nevertheless, one can easily show, via the invariance of the trace $\text{Tr}\hat A\hat B=\text{Tr}\hat B\hat A$, that $G(\beta^-,\tau')=-G(0^+,\tau')$, when $0<\tau'<\beta$, and $G(\tau,\beta^-)=-G(\tau,0^+)$, when $0<\tau<\beta$. This means that we can still expand the time dependence in terms of a Matsubara-frequency matrix, with a separate Matsubara-frequency expansion for the $\tau$ and $\tau'$ dependences, respectively. This yields
\begin{equation}
\label{gmat}
    G(\tau,\tau')=T\sum_m\sum_n\re^{-\ri\omega_m\tau+\ri\omega_n\tau'}G_{mn},
\end{equation}
where $\ri\omega_m=2\ri\piup T\left (m+{1}/{2}\right )$ is the fermionic Matsubara frequency, for $m \in\mathbb{Z}$. The Green's function in Matsubara frequency space is expressed as 
\begin{equation}
\label{Gftdef}
    G_{mn}=T\int_0^{\beta}\,\rd\tau\ \int_0^{\beta}\,\rd\tau'\ \re^{\ri\omega_m\tau-\ri\omega_n\tau'}G(\tau,\tau')
\end{equation} 
and the dynamical mean-field is given by a similar
expression
\begin{equation}
    \lambda_{mn}=T\int_0^{\beta}\,\rd\tau\ \int_0^{\beta}\,\rd\tau'\ \re^{\ri\omega_m\tau-\ri\omega_n\tau'}\lambda(\tau,\tau').
\end{equation}
Note that we are working with an imaginary-time formalism here. When one is in equilibrium, with a time-independent Hamiltonian (no external field), the Green's function and the dynamical mean-field become time-translation invariant, and the Matsubara-frequency representation of the Green's function and the dynamical mean-field become \textit{diagonal} matrices.
Finally, we define the frequency-dependent charge susceptibility as 
\begin{equation}
\label{chidef}
    \chi(\ri\nu_{\el}) = T\int_0^{\beta}\,\rd\tau\ \left\{\langle \hat{n}(\tau)\hat{n}(0)\rangle -\langle\hat{n}(\tau)\rangle\langle\hat{n}(0)\rangle\right\}\re^{\inul\tau},
\end{equation}
where $\inul$ is a bosonic Matsubara frequency, $\inul=2\ri\piup \ell T$, for $\ell\in\mathbb{Z}$. Next, note that $[\hat{\mathcal{H}},\hat n(\tau)]=0$, so that the second term of the susceptibility in equation~(\ref{chidef}) gives a $\delta_{\ell,0}$ when integrating over $\tau$ (a similar result holds for $q=0$ in the lattice problem, since the total electron number is conserved). One might have thought that the first term also has no imaginary time dependence, but when $\tau\ne0$, the expectation value has one $\hat n$ operator located at $\tau=0$ and one located at $\tau$; the expectation value has an intrinsic $\tau$ dependence inherited from the dynamical mean-field due to the time ordering. We are interested in the non-zero frequency response, so dropping the zero frequency component, our expression for the susceptibility becomes 
\begin{equation}
\label{susceptibility}
    \chi(\inul)=T\int_0^{\beta}\,\rd\tau\langle\hat{n}(\tau)\hat{n}(0)\rangle \re^{\inul\tau}.
\end{equation}

To find the equation which governs the frequency-dependent susceptibility, we introduce a time-dependent field $h(\tau)$ into our impurity Hamiltonian,
\begin{equation}
    \hat{\mathcal{H}}'(\tau)=\hat{\mathcal{H}}-h(\tau)\hat{n}
\end{equation}
and take the limit $h(\tau)\to 0$ at the end of the calculation. This is equivalent to replacing the dynamical mean-field $\lambda$ in the evolution operator with 
\begin{equation}
    \lambda(\tau,\tau')\to \lambda(\tau,\tau')-h(\tau)\delta(\tau-\tau')
\end{equation}
so that the evolution operator in equation~(\ref{timev}) becomes 
\begin{equation}
\label{timev2}
    S(\lambda,h)=\exp\left \{-\int_0^{\beta}\,\rd\tau\ \int_0^{\beta}\,\rd\tau'\ \hat{c}^{\dagger}(\tau)\big[\lambda(\tau,\tau')-h(\tau)\delta(\tau-\tau')\big]\hat{c}(\tau')\right \}
\end{equation}
and the Green's function satisfies 
\begin{equation}
\label{gf}
    G(\tau,\tau')=-\frac{1}{\mathcal Z}\text{Tr}\left\{\re^{-\beta \hat{\mathcal H}}\mathcal{T}_{\tau}\hat{S}(\lambda,h)\hat{c}(\tau)\hat{c}^{\dagger}(\tau')\right\}.
\end{equation}
In this fashion, we find it more convenient to think of the $h$-field as being added to the dynamical mean-field.
It is important to note that the introduction of the $h$-field breaks the time-translation invariance of the problem.
Expressing the auxiliary field $h(\tau,\tau')$ as a matrix in Fourier space, we have 
\begin{equation}
    h_{mn}=T\int_0^{\beta}\,\rd\tau\ \int_0^{\beta}\,\rd\tau' \re^{\ri\omega_m\tau-\ri\omega_n\tau'}\delta(\tau-\tau')h(\tau)
    =T\int_0^{\beta}\,\rd\tau\ \re^{(\ri\omega_m-\ri\omega_n)\tau}h(\tau)
    =h_{m-n}\,.
\end{equation}
This means that, in Matsubara space, the auxiliary field takes the form of a Toeplitz matrix, $h_{mn}=h_{m-n}$; a Toeplitz matrix has all the elements on each subdiagonal equal to each other (but they can vary for different subdiagonals). Note that the difference in fermionic Matsubara frequencies is a bosonic Matsubara frequency, so we define the auxiliary field $h(\tau)$ as a sum over bosonic Matsubara frequencies, 
\begin{equation}
\label{hmats}
    h(\tau)=T\sum_{\ell}\re^{\ri\nu_{\ell}\tau}h_{\ell}
\end{equation}
and by inversion we have
\begin{equation}
    h_{\ell}=\int_0^{\beta}\,\rd\tau\ \re^{-\ri\nu_{\ell}\tau}h(\tau).
\end{equation}
By comparison with the matrix form $h_{mn}=h_{m-n}$, we see that the matrix elements are determined by bosonic Matusbara frequencies, $\ri\nu_{\el}=\ri\omega_{n}-\ri\omega_{m}$.
We rewrite the susceptibility as a functional derivative of the Green's function, so that equation~(\ref{susceptibility}) becomes 
\begin{equation}
    \chi(\inul)=\lim_{h\to 0}\,T\sum_{m,n}\frac{\partial G_{mn}}{\partial h_{\ell}}\,.
\end{equation}
Note that {\it after} the derivative is taken, and $h\to 0$, the Green's functions in the final expression have their time-translation invariance restored. \\The above equation provides an explicit expression for the susceptibility in terms of a functional derivative of the Green's function. In cases where we can \textit{explicitly} determine how $G_{mn}$ depends on the field $h_{\ell}$  (for small fields), this equation presents us with an alternative method to determine the dynamic susceptibility without solving the Bethe-Salpeter equation. The equivalence of this and the Bethe-Salpeter equation derived below will be demonstrated in the context of the Falicov-Kimball model, where such an explicit relation can be found. 

\section{Derivation of the frequency-dependent susceptibility}
\subsection{Determination of the Green's function to linear order in the perturbing field}
To begin, we consider the Green's function for a system with an impurity Hamiltonian $\hat{\mathcal{H}}$ in the presence of a (time-translation invariant) dynamical mean-field $\lambda(\tau-\tau')$. The full Hamiltonian is given by \begin{equation}
    \hat{\mathcal{H}}'(\tau)=\hat{\mathcal{H}}-h(\tau)\hat{n}.
\end{equation}
As stated above, this is equivalent to the system evolving via the unperturbed Hamiltonian $\hat{\mathcal{H}}$ in the presence of a new dynamical mean-field, \begin{equation}
    \lambda(\tau-\tau')\to \lambda(\tau-\tau')-h(\tau)\delta(\tau-\tau').
\end{equation}
Since the time-ordering operator allows us to freely manipulate the exponentials, we find that
\begin{eqnarray}
     S(\lambda,h)&=&\mathcal{T}_\tau\exp\left \{-\int_0^{\beta}\,\rd\tau\ \int_0^{\beta}\,\rd\tau'\ \hat{c}^{\dagger}(\tau)\big[\lambda(\tau-\tau')-h(\tau)\delta(\tau-\tau')\big]\hat{c}(\tau')\right \} \\
     &=&\mathcal{T}_\tau\left [S(\lambda)\exp\left\{-\int_0^{\beta}\,\rd\tau\ \int_0^{\beta}\,\rd\tau'\ \hat{c}^{\dagger}(\tau)h(\tau)\delta(\tau-\tau')\hat{c}(\tau')\right\}\right ]
\\
    &=&\mathcal{T}_\tau\left [S(\lambda)\exp\left\{-\int_0^{\beta}\,\rd\tau\ h(\tau)\hat{n}(\tau)\right\}\right ].
\end{eqnarray}
We expand the exponential of the external field to linear order in the field $h(\tau)$, \begin{equation}
    \exp\left\{-\int_0^{\beta}\,\rd\tau\ h(\tau)\hat{n}(\tau)\right\} \approx 1 -\int_0^{\beta}\,\rd\tau\ h(\tau)\hat{n}(\tau),
\end{equation}
then the equation for the Green's function, to linear order in the presence of the $h$-field, becomes%
\begin{eqnarray}
    G(\tau,\tau')
    &\approx&-\frac{1}{\mathcal Z}\text{Tr}\left\{\re^{-\beta \hat{\mathcal H}}\mathcal{T}_{\tau}\hat{S}(\lambda)\hat{c}(\tau)\hat{c}^{\dagger}(\tau')\right\}+\frac{1}{\mathcal Z}\int_0^{\beta}\,\rd\bar{\tau}\ h(\bar{\tau})\text{Tr}\left\{\re^{-\beta \hat{\mathcal H}}\mathcal{T}_{\tau}\hat{S}(\lambda)n(\bar{\tau})\hat{c}(\tau)\hat{c}^{\dagger}(\tau')\right\}\nonumber\\
    &&{}+\mathcal{O}(h^2).
\end{eqnarray}
The first term is the equilibrium Green's function.  Expanding the perturbing field $h(\tau)$ in Matsubara frequencies, and assuming a pure tone (only one nonzero Matsubara frequency component to the external field) the above becomes 
\begin{equation}
    G(\tau,\tau')=G_{\rm eq}(\tau-\tau')+\int_0^{\beta}\,\rd\bar{\tau}\ h_{\el}\re^{\ri\nu_{\el}\bar{\tau}}\left\langle \mathcal{T}_{\tau}\big[\hat{n}(\bar{\tau})\hat{c}(\tau)\hat{c}^{\dagger}(\tau')\big]\right\rangle.
\end{equation}
Since the object \begin{equation}
\left    \langle \mathcal{T}_{\tau}\big[\hat{n}(\bar{\tau})\hat{c}(\tau)\hat{c}^{\dagger}(\tau')\big]\right\rangle
\end{equation}
is governed only by the evolution operator $S(\lambda)$ and not the perturbing field $h$, it is an equilibrium correlation function and must be time-translation invariant. Hence, it only depends on the time differences $\tau-\tau', \tau-\bar{\tau}, \tau'-\bar{\tau}$. We now Fourier transform $G(\tau,\tau')$ to Matsubara space via equation~(\ref{Gftdef}), with the result %
\begin{equation}
    G_{mn}=\delta_{mn}G_m+\int_0^{\beta}\,\rd\tau\ \int_0^{\beta}\,\rd\tau'\ \int_0^{\beta}\,\rd\bar{\tau}\ h_{\el}\re^{\ri\omega_m\tau-\ri\omega_n\tau'+\ri\nu_{\el}\bar{\tau}}\left\langle \mathcal{T}_{\tau}\big[\hat{n}(\bar{\tau})\hat{c}(\tau)\hat{c}^{\dagger}(\tau')\big]\right\rangle.
\end{equation}
Since the correlation function depends only on time differences, shifting $\tau\to\tau+\bar{\tau}, \tau'\to\tau'+\bar{\tau}$ removes the $\bar{\tau}$ dependence from the correlation function. We are then able to perform the integral over $\bar{\tau}$ which results in a delta function $\delta(\ri\omega_m-\ri\omega_n+\ri\nu_{\el})$, forcing $\ri\omega_n=\ri\omega_m+\ri\nu_{\el}$. Therefore, the only non-vanishing contribution to $G_{mn}$ which is proportional to $h_{\el}$ is along the diagonal $m, m+\el$. This result is critical to our derivation of a Bethe-Salpeter-like equation below.

\subsection{Derivation of a Bethe-Salpeter-like Equation}
In Fourier space, the matrix inverse of $G$ is found from the equation of motion, and is
given by 
\begin{equation}
    G_{mn}^{-1}=(\ri\omega_m+\mu)\delta_{mn}-\lambda_{mn}+h_{m-n}-\Sigma_{mn}\,.
    \label{eq: g_inv}
\end{equation}
This result holds in equilibrium {\it and} nonequilibrium (of course, in equilibrium, we have $h=0$ and both $\lambda$ and $\Sigma$ are proportional to $\delta_{mn}$).
Next, we rewrite the matrix $G$ schematically as the matrix multiplication $GG^{-1}G$ in the expression for the susceptibility. This yields
\begin{equation}
    \chi(\inul)=\lim_{h\to 0}\,T\sum_{mn}\sum_{rs}\frac{\partial}{\partial h_{\el}}\left[G_{mr}G^{-1}_{rs}G_{sn}\right]
    =2\chi(\inul)+\lim_{h\to 0}T\sum_{mn}\sum_{rs}G_{mr}\frac{\partial G^{-1}_{rs}}{\partial h_{\el}}G_{sn}
\end{equation}
so 
\begin{equation}
    \chi(\inul)=-\lim_{h\to 0}\,T\sum_{mn}\sum_{rs}G_{mr}\frac{\partial G_{rs}^{-1}}{\partial h_{\el}}G_{sn}
    =-\lim_{h\to 0}T\sum_{mn}\sum_{rs}G_{mr}\left[\delta_{r-s,\ell}-\frac{\partial\Sigma_{rs}}{\partial h_{\ell}}\right ]G_{sn}\,,
        \label{latest}
\end{equation}
where the derivative of $G^{-1}$ with respect to $h_{\ell}$ is found explicitly from the field term in equation~(\ref{eq: g_inv}) and implicitly from the dependence of the self-energy on the field.
After the derivative has been performed, we take the limit $h\to 0$, which restores time-translation invariance to the Green's function, given by $G_{mn}\to G_m\delta_{mn}$. Taking this limit where appropriate and simplifying gives 
\begin{equation}
    \chi(\inul)=-T\sum_{n}G_nG_{n+\el}+T\sum_{mn}G_mG_n\frac{\partial \Sigma_{mn}}{\partial h_{\el}}\,.
\end{equation}
Using the fact that the self-energy is a functional of the Green's function, we employ the chain rule yielding
\begin{equation}
    \chi(\inul)=-T\sum_{n}G_nG_{n+\el}+T\sum_{mn}\sum_{m'n'}G_mG_n\frac{\partial \Sigma_{mn}}{\partial G_{m'n'}}\frac{\partial G_{m'n'}}{\partial h_{\el}}\,.
\end{equation}

The functional derivative of the self-energy with respect to the Green's function (which is proportional to the irreducible vertex function) does not have all four frequency indices independent of each other. To see this, we
write the functional derivative ${\partial\Sigma_{mn}}/{\partial G_{m'n'}}$ in terms of integrals over imaginary time, 
\begin{equation}
\label{timedep}
    \frac{\partial\Sigma_{mn}}{\partial G_{m'n'}}=T^4\int_0^{\beta}\,\rd\tau_1\int_0^{\beta} \,\rd\tau_2\int_0^{\beta}\,\rd\tau_1'\int_0^{\beta} \,\rd\tau_2'\ \exp\left[\ri\omega_m\tau_1-\ri\omega_n\tau_2-\ri\omega_{m'}\tau_1'+\ri\omega_{n'}\tau_2'\right]\frac{\partial\Sigma(\tau_1,\tau_2)}{\partial G(\tau_1',\tau_2')}\,,
\end{equation}
and we recall that we must take the limit $h\to 0$ (after taking the functional derivative). When we do, the problem becomes time-translation invariant. Hence,%
\[
\lim_{h\to 0}\frac{\partial\Sigma(\tau_1,\tau_2)}{\partial G(\tau_1',\tau_2')}
\] 
\emph{must be independent of absolute time}, depending only on relative times $\tau_1-\tau_2$,  $\tau_1'-\tau_2'$ and $\tau_1-\tau_1'$. Consequently, we can shift the times in equation~(\ref{timedep}), performing the change of variables $\tau_1\to\tau_1+\tau_2',$ $\tau_2\to\tau_2+\tau_2',$
    $\tau_1'\to\tau_1'+\tau_2'$.
Then, 
\[
\lim_{h\to 0}\frac{\partial\Sigma(\tau_1,\tau_2)}{\partial G(\tau_1',\tau_2')}
\] 
becomes independent of $\tau_2'$ and performing the integral over $\tau_2'$ results in a delta function $\delta(\ri\omega_m+\ri\omega_{n'}-\ri\omega_n-\ri\omega_{m'})$,
forcing $\ri\omega_n= \ri\omega_m+\ri\omega_{n'}-\ri\omega_{m'}$.
Consequently, the susceptibility becomes 
\begin{equation}
\label{latest2}
    \chi(\inul)=-T\sum_nG_{n}G_{n+\el}+T\sum_{m,n',m'}G_{m}G_{m+n'-m'}\frac{\partial\Sigma_{m,m+n'-m'}}{\partial G_{m'n'}}\frac{\partial G_{m'n'}}{\partial h_{\el}}\,.
\end{equation}
Note that one cannot still assume time-translation invariance for the $G_{m'n'}$ in the summation above, until after we have taken the derivative with respect to $h_l$.

We are only capturing the linear response to the imposed external field.
Since the dynamic susceptibility $\chi(\inul)$ depends explicitly on %
\[
\frac{\partial G_{mn}}{\partial h_{\ell}}\big\vert_{h_{\ell}\to 0}\,,
\]
  only the contributions to $G_{mn}$ proportional to $h_{\ell}$ survive. This means that, without loss of generality, we again consider the external field to have just one single Fourier component, given by $h_{\ell}$ with $\ell\ne 0$. As demonstrated in the section above, the only elements of $G_{mn}$ which are linear in $h_{\ell}$ are those for which $G_{mn}=G_{m,m+\ell}$.
Hence, revisiting equation~(\ref{latest2}), only elements with $\ri\omega_{n'}-\ri\omega_{m'}=\ri\nu_{\el}$ are nonzero. Therefore, %
\begin{equation}
    \chi(\inul)=-T\sum_nG_nG_{n+\el}+T\sum_{m,m'}G_mG_{m+\el}\frac{\partial\Sigma_{m,m+\el}}{\partial G_{m',m'+\el}}\frac{\partial G_{m',m'+\el}}{\partial h_{\el}}\,.
\end{equation}
We make the definitions 
\begin{eqnarray}
    \chi_0(\ri\omega_n;\ri\nu_\el)&=&-G_nG_{n+\el}\,,
\\[2ex]
    \chi(\ri\omega_m,\ri\omega_{n};\ri\nu_\el)&=&\frac{\partial G_{mn}}{\partial h_\el}
\end{eqnarray}
and
\begin{equation}
    \Gamma(\ri\omega_m,\ri\omega_{m'};\ri\nu_\el)\hspace{1mm}=\frac{1}{T}\frac{\partial\Sigma_{m,m+\el}}{\partial G_{m',m'+\el}}
    \label{eq: gamma_def}
\end{equation}
then, we have 
\begin{equation}
    \chi(\inul)=T\sum_n\chi_0(i\omega_n;i\nu_\el)-T^2\sum_{m,m'}\chi_0(i\omega_m;i\nu_\el)\Gamma(i\omega_m,i\omega_{m'};i\nu_\el)\chi(i\omega_{m'},i\omega_{m'+\el};i\nu_\el).
\end{equation}
Unlike the conventional Bethe-Salpeter equation, written in terms of matrices, we find an equation that has a vector character to it and is given by
\begin{equation}
\label{final}
  \chi(\ri\omega_m,\ri\omega_{m+\el};\ri\nu_\el)=\chi_0(\ri\omega_m;i\nu_\el)-\chi_0(\ri\omega_m;\ri\nu_\el)T\sum_{m'}\Gamma(\ri\omega_m,\ri\omega_{m'};\ri\nu_\el)\chi(\ri\omega_{m'},\ri\omega_{m'+\el};\ri\nu_\el)
\end{equation}
with the susceptibility determined from
\begin{equation}
\label{final2}
    \chi(\inul)=T\sum_{m}\chi(\ri\omega_m,\ri\omega_{m+\el};\ri\nu_\el).
\end{equation}
This vector character can be viewed as the dressed or retarded interaction between a particle-hole pair and the exchanged boson mediating the interaction, as is commonly used in triangular vertices~\cite{parcollet}.

The generalization to the lattice is more complicated, because of the additional momentum dependence of the external field, which introduces an additional momentum dependence to the susceptibilities. We do not discuss those issues further here.

\section{Application to the Falicov-Kimball Model}
For the Falicov-Kimball model, introduced in \cite{fk}, the impurity Hamiltonian takes the form $\hat{\mathcal{H}}=\hat{\mathcal{H}}_0+\hat{\mathcal{V}}$, where
\begin{equation}
    \hat{\mathcal{H}}_0=-\mu\hat{n}~~\text{and}~~\hat{\mathcal{V}}=Uw_1\hat n
\end{equation}
with $w_1$ a classical variable equal to 0 or 1 depending on whether a heavy fermion is present or not and the impurity is placed
in the presence of the dynamical mean-field $\lambda(\tau-\tau')$.

In equilibrium, the Green's function can be solved in terms of the expectation value of $w_1$, denoted $\langle w_1\rangle$, which is the heavy fermion density on the impurity. We find that the equilibrium Falicov-Kimball model Green's function is
\begin{equation}
    G_{mn}^{\mathrm{FK,eq}}=\left (\frac{1-\langle w_1\rangle}{\ri\omega_m+\mu-\lambda_m}+\frac{\langle w_1\rangle}{\ri\omega_m+\mu-U-\lambda_m}\right )\delta_{mn}=G_m^{\mathrm{FK,eq}}\delta_{mn}
\end{equation}
and the equilibrum self-energy is defined from the Dyson equation $( G_{mn}^{\mathrm{FK,eq}})^{-1}=(\ri\omega_m+\mu-\lambda_m)\delta_{mn}-\Sigma_{mn}^{\mathrm{FK,eq}}$, which becomes
\begin{equation}
\Sigma_{mn}^{\mathrm{FK,eq}}=\left (U\langle w_1\rangle +\frac{\langle w_1\rangle(1-\langle w_1\rangle)U^2}{\ri\omega_m+\mu-(1-\langle w_1\rangle)U-\lambda_m}\right )\delta_{mn}=\Sigma_m^{\mathrm{FK,eq}}\delta_{mn}\,.
\end{equation}
This Falicov-Kimball model Green's function can also be expressed as a weighted sum of the Green's function for the impurity and the Green's function for the impurity with $\mu\to\mu-U$. This is written as
\begin{equation}
    G_{mn}^{\mathrm{FK,eq}}=\left [(1-\langle w_1\rangle)G_{mn}+\langle w_1\rangle G_{mn}\Big |_{\mu\to\mu-U}\right ].
\end{equation}
This latter form also holds in nonequilibrium.

\subsection{The Falicov-Kimball Vertex}

We obtain the vertex function for the Falicov-Kimball model without defining an ``auxillary Green's function'' as in \cite{Jim1} or using diagrammatic techniques as in \cite{andrij}. We begin with the Green's function for a spinless fermion with Hamiltonian $\hat{\mathcal{H}}_0=-\mu\hat{n}$ in the dynamical mean-field plus the external field, 
\begin{eqnarray}
    G(\tau,\tau')&=&-\frac{1}{\mathcal{Z}}\theta(\tau-\tau')\text{Tr}\left\{\mathcal{T}_{\tau}\re^{-\beta \hat{\mathcal{H}_0}}\mathcal{T}_{\tau}S(\lambda,h)\hat{c}(\tau)\hat{c}^{\dagger}(\tau')\right\}\nonumber\\
    &&{}+\frac{1}{\mathcal{Z}}\theta(\tau'-\tau)\text{Tr}\left\{\mathcal{T}_{\tau}\re^{-\beta \hat{\mathcal{H}_0}}\mathcal{T}_{\tau}S(\lambda,h)\hat{c}^{\dagger}(\tau')\hat{c}(\tau)\right\}.
\end{eqnarray}
Using the equation of motion and expressing it in Matsubara space, we obtain 
\begin{equation}
    (\ri\omega_m+\mu-\lambda_m)G_{mn}+Th_{\el}G_{m+\el n}=\delta_{mn}\,.
\end{equation}
This is recast as a matrix equation
\begin{equation}
    \sum_{m'} \left[(\ri\omega_m+\mu-\lambda_m)\delta_{mm'}+Th_{\el}\delta_{m,m'-l}\right]G_{m'n}=\delta_{mn}\,,
\end{equation}
so we can identify the quantity in square brackets as the inverse Green's function of the impurity, $G^{-1}$, which is given by
\begin{equation}
    G_{mn}^{-1}=(\ri\omega_m+\mu-\lambda_m)\delta_{mn}+Th_{\el}\delta_{m,n-\el}\,.
\end{equation}
We want to compute the Green's function as the inverse of $G^{-1}$ to lowest order in the external field $h_\el$.
We re-arrange the equation as follows,
\begin{equation}
   G^{-1}_{mn} =\sum_{m'}(\ri\omega_m+\mu-\lambda_m)\delta_{mm'}\left[\delta_{m'n}+\frac{Th_{\el}}{\ri\omega_{m'}+\mu-\lambda_{m'}}\delta_{m',n-\el}\right]
\end{equation}
and use the fact that $h_l$ is small to allow us to invert this matrix inverse  via the geometric series and obtain $G_{mn}$ of the impurity as 
\begin{equation}
    G_{mn}=\frac{\delta_{mn}}{\ri\omega_m+\mu-\lambda_m}-\frac{Th_{\el}\delta_{m,n-\el}}{(\ri\omega_m+\mu-\lambda_m)(\ri\omega_n+\mu-\lambda_n)}+\mathcal{O}(h_{\el}^2)
\end{equation}
to lowest order in the external field. Note that the first term is the equilibrium result and the second term is the lowest-order correction due to the external field.

The results for the Falicov-Kimball model then follow by a weighted sum of the impurity results for empty and occupied heavy electrons on the impurity site (just as we identified in equilibrium). This yields 
\begin{equation}
    G_{mn}^{\mathrm{FK}}=(1-\langle w_1\rangle )G_{mn}+\langle w_1\rangle G_{mn}\big\vert_{\mu\to\mu-U}\,.
\end{equation}
Thus, to order $h_{\el}$, we have 
\begin{eqnarray}
    G_{mn}^{\mathrm{FK}}&=&(1-\langle w_1\rangle)\bigg[\frac{\delta_{mn}}{\ri\omega_m+\mu-\lambda_m}-\frac{Th_{\el}\delta_{m,n-\el}}{(\ri\omega_m+\mu-\lambda_m)(\ri\omega_n+\mu-\lambda_n)}\bigg]\nonumber\\[2ex]
  &&{}  +\langle w_1\rangle \bigg[\frac{\delta_{mn}}{\ri\omega_m+\mu-U-\lambda_m}-\frac{Th_{\el}\delta_{m,n-\el}}{(\ri\omega_m+\mu-U-\lambda_m)(i\omega_n+\mu-U-\lambda_n)}\bigg]
\\[3ex]
    &=&G_m^{\mathrm{FK,eq}}\delta_{mn}-Th_{\el}\delta_{m,n-\el}\bigg[\frac{1-\langle w_1\rangle}{(\ri\omega_m+\mu-\lambda_m)(\ri\omega_n+\mu-\lambda_n)}\nonumber\\[2ex]
    &&\qquad\qquad \,\, {}+\frac{\langle w_1\rangle}{(\ri\omega_m+\mu-U-\lambda_m)(\ri\omega_n+\mu-U-\lambda_n)}\bigg],
\end{eqnarray}
where again, we separate with respect to the equilibrium result and the lowest-order corrections due to the external field.
We denote the lowest-order correction via $\delta G_{mn}^{\mathrm{FK}}$ and after putting everything over a common denominator, we find 
\begin{equation}
\label{deltaG}
    \!\delta G_{mn}^{\mathrm{FK}}\!=\!-Th_{\el}\delta_{m,n-\el}\!\bigg[\!\frac{(\ri\omega_m+\mu-\lambda_m)(\ri\omega_n+\mu-\lambda_n)-U(1-\langle w_1\rangle)(\ri\omega_m+\mu-\lambda_m+\ri\omega_n+\mu-\lambda_n-U)}{(\ri\omega_m+\mu-\lambda_m)(\ri\omega_n+\mu-\lambda_n)(\ri\omega_m+\mu-U-\lambda_m)(\ri\omega_n+\mu-U-\lambda_n)}\!\bigg].
    \vphantom{\frac{\frac12^1}{\frac12^1}}
    \end{equation}
To calculate the vertex, we also need to determine the lowest-order change to the self-energy, denoted by $\delta \Sigma_{mn}^{\mathrm{FK}}$. The total self-energy is defined from the Dyson equation 
\begin{equation}
    (G_{mn}^{\mathrm{FK}})^{-1}=(\ri\omega_m+\mu-\lambda_m)\delta_{mn}+Th_{\el}\delta_{m,n-\el}-\Sigma_{mn}^{\mathrm{FK}}\,.
\end{equation}
We separate out the equilibrium and lowest-order corrections due to the field via $\Sigma_{mn}^{\mathrm{FK}}=\Sigma_m^{\mathrm{FK,eq}}\delta_{mn}+\delta\Sigma_{mn}^{\mathrm{FK}}$, then factor the Falicov-Kimball model Green's function factor out to the left 
\[
\left[(G^{\mathrm{FK,eq}}_m)^{-1}=\ri\omega_m+\mu-\lambda_m-\Sigma_m^{\mathrm{FK,eq}}\right]
\] 
to yield
\begin{equation}
    \left(G_{mn}^{\mathrm{FK}}\right)^{-1}=\sum_{m'}\left(G_m^{\mathrm{FK,eq}}\right)^{-1}\delta_{mm'}\left[\delta_{m'n}+G^{\mathrm{FK,eq}}_{m'}\left(Th_{\el}\delta_{m',n-\el}-\delta\Sigma_{m'n}^{\mathrm{FK}}\right)\right].
\end{equation}
From this we can rewrite $G_{mn}^{\mathrm{FK}}$ (again to order $h_{\el}$) as \begin{equation}
    G_{mn}^{\mathrm{FK}}=G_m^{\mathrm{FK,eq}}\delta_{mn}-Th_{\el}G_m^{\mathrm{FK,eq}}\delta_{m,n-\el}G_n^{\mathrm{FK,eq}}+G_m^{\mathrm{FK,eq}}\delta\Sigma_{mn}^{\mathrm{FK}}G_n^{\mathrm{FK,eq}}
\end{equation}
so that \begin{equation}
    \delta G_{mn}^{\mathrm{FK}}=-Th_{\el}G_m^{\mathrm{FK,eq}}G_n^{\mathrm{FK,eq}}\delta_{m,n-\el}+G_m^{\mathrm{FK,eq}}\delta\Sigma_{mn}^{\mathrm{FK}}G_n^{\mathrm{FK,eq}}
\end{equation}
 consequently, 
\begin{equation}
    \delta\Sigma_{mn}^{\mathrm{FK}}=\left(G_m^{\mathrm{FK,eq}}\right)^{-1}\delta G_{mn}^{\mathrm{FK}}\left(G_n^{\mathrm{FK,eq}}\right)^{-1}-Th_{\el}\delta_{m,n-\el}\,.
\end{equation}
By substituting in $\delta G_{mn}^{\mathrm{FK}}$ from equation~(\ref{deltaG}) and after some significant algebra, we find 
\begin{align}
&    \delta\Sigma_{mn}^{\mathrm{FK}}=\delta G_{mn}^{\mathrm{FK}}\nonumber\\    
    &\times\frac{U^2 \langle w_1\rangle (1-\langle w_1\rangle)}{G_m^{\mathrm{FK,eq}}G_n^{\mathrm{FK,eq}}\left [\big(\ri\omega_m+\mu-\lambda_m-(1-\langle w_1\rangle)U\big)\big(\ri\omega_n+\mu-\lambda_n-(1-\langle w_1\rangle)U\big)+\langle w_1\rangle(1-\langle w_1\rangle)U^2\right ]}\,.\nonumber\\
\end{align}
Finally, by computing the differences $G_m^{\mathrm{FK,eq}}-G_n^{\mathrm{FK,eq}}$, $\Sigma_m^{\mathrm{FK,eq}}-\Sigma_n^{\mathrm{FK,eq}}$ of the equilibrium Green's function and self-energies, we can replace the complicated expression with a much simpler one given by
\begin{align}
    &\frac{\Sigma_m^{\mathrm{FK,eq}}-\Sigma_n^{\mathrm{FK,eq}}}{G_m^{\mathrm{FK,eq}}-G_n^{\mathrm{FK,eq}}}={}\nonumber\\
    &=\frac{U^2\langle w_1\rangle(1-\langle w_1\rangle)}{G_m^{\mathrm{FK,eq}}G_n^{\mathrm{FK,eq}}\left [\big(\ri\omega_m+\mu-\lambda_m-(1-\langle w_1\rangle)U\big)\big(\ri\omega_n+\mu-\lambda_n-(1-\langle w_1\rangle)U\big)+\langle w_1\rangle(1-\langle w_1\rangle)U^2\right ]}\,.\nonumber\\
\end{align}
Therefore, we have 
\begin{equation}
    \delta\Sigma_{mn}^{\mathrm{FK}}=\delta G_{mn}^{\mathrm{FK}}\frac{\Sigma_m^{\mathrm{FK,eq}}-\Sigma_n^{\mathrm{FK,eq}}}{G_m^{\mathrm{FK,eq}}-G_n^{\mathrm{FK,eq}}}\,.
\end{equation}
Hence, the irreducible vertex for the Falicov-Kimball model, defined in equation~(\ref{eq: gamma_def}), becomes
\begin{equation} \Gamma^{\mathrm{FK}}(\ri\omega_m,\ri\omega_{m'};\ri\nu_\el)=\frac{1}{T}\frac{\delta\Sigma_{mm+l}^{\mathrm{FK}}}{\delta G_{m'm'+\el}^{\mathrm{FK}}}\,.
\end{equation}
Thus, we see that for the Falicov-Kimball model, the vertex function is diagonal and takes the simple form
\begin{align}
    \Gamma^{\mathrm{FK}}(\ri\omega_m,\ri\omega_{m'};\ri\nu_\el)=\delta_{m,m'}\frac{1}{T}\frac{\Sigma_m^{\mathrm{FK,eq}}-\Sigma_{m+\el}^{\mathrm{FK,eq}}}{G_{m}^{\mathrm{FK,eq}}-G_{m+\el}^{\mathrm{FK,eq}}}\,.
\end{align}
While the algebra is a bit messy and long, we feel this derivation is simpler than  previous derivations.

\subsection{Comparison of Susceptibilities for the Falicov-Kimball Model}

The perturbative approach we follow here provides us with an alternative way to directly calculate the susceptibility without needing the Bethe-Salpeter-like equation in equation~(\ref{final}). Recall that
the susceptibility is given by 
\begin{equation}
    \chi(\ri\nu_{\el})=T\sum_{m}\frac{\delta G_{mm+l}}{\delta h_{\el}}
\end{equation}
for $\el\neq 0$. From equation~(\ref{deltaG}), we see that the derivative of $G^{\mathrm{FK}}$ with respect to $h_\el$ is computed directly from the coefficient of $h_{\el}$ in $\delta G_{mn}^{\mathrm{FK}}$. Thus, the dynamic susceptibility becomes 
\begin{align}
    &\chi^{\mathrm{FK}}(\ri\nu_{\el})={}\nonumber\\
    &=-T\sum_{m}\frac{(\ri\omega_m+\mu-\lambda_m)(\ri\omega_{m+\el}+\mu-\lambda_{m+\el})-U(1-\langle w_1\rangle)(\ri\omega_m+\ri\omega_{m+\el}+2\mu-\lambda_m-\lambda_{m+\el}-U)}{(\ri\omega_m+\mu-\lambda_m)(\ri\omega_{m+\el}+\mu-\lambda_{m+\el})(\ri\omega_m+\mu-U-\lambda_m)(\ri\omega_{m+\el}+\mu-U-\lambda_{m+\el})}\,.\nonumber\\
\end{align}
We want to show that this result is the same as what we find from the Bethe-Salpeter-like equation. 
Using the fact that the Falicov-Kimball vertex is diagonal (proportional to $\delta_{m,m'}$), we solve equation~(\ref{final2}) exactly for $\chi(\ri\nu_{\el})$ with the result  \begin{align}
    \chi^{\mathrm{FK}}(\ri\nu_{\el})=T\sum_m\frac{\chi_0(\ri\omega_m;\ri\nu_\el)}{1+\chi_0(\ri\omega_m;\ri\nu_\el)
    \frac{\Sigma_m^{\mathrm{FK,eq}}-\Sigma_{m+\el}^{\mathrm{FK,eq}}}{G_m^{\mathrm{FK,eq}}-G_{m+\el}^{\mathrm{FK,eq}}}}\,.
\end{align}
Substituting in $\chi_0(\ri\omega_m;\ri\nu_\el)=-G_m^{\mathrm{FK,eq}}G_{m+l}^{\mathrm{FK,eq}}$ and after some lengthy algebra, we arrive at the same result as above, namely
\begin{align}
    &\chi^{\mathrm{FK}}(\ri\nu_{\el})\nonumber\\
    &=-T\sum_{m}\frac{(\ri\omega_m+\mu-\lambda_m)(\ri\omega_{m+\el}+\mu-\lambda_{m+\el})-U(1-\langle w_1\rangle)(\ri\omega_m+\ri\omega_{m+\el}+2\mu-\lambda_m-\lambda_{m+\el}-U)}{(\ri\omega_m+\mu-\lambda_m)(\ri\omega_{m+\el}+\mu-\lambda_{m+\el})(\ri\omega_m+\mu-U-\lambda_m)(\ri\omega_{m+\el}+\mu-U-\lambda_{m+\el})}.\nonumber\\
\end{align}
This verifies that if we have an expression for $\delta G_{mn}$ in terms of $h_{\el}$, taking the functional derivative is a much simpler way to determine the dynamical susceptibility. 

\section{Conclusion}

This paper is primarily pedagogical. It shows a straightforward, simple, and direct methodology for calculating dynamical susceptibilities. We show how the conventional Bethe-Salpeter (matrix) equation can be simplified
to a vector-like equation for systems that can be mapped to an impurity Hamiltonian, and hence treated via dynamical mean-field theory (DMFT). The resulting vector equation allows for an immediate solution when the vertex function is diagonal (as it occurs in the Falicov-Kimball model). In addition, in cases where  one can determine the functional relationship of the nonequilibrium Green's function with respect to the applied field, the dynamic susceptibility is computed directly avoiding the need for the Bethe-Salpeter-like equation altogether. We also show a new derivation for the vertex function of the Falicov-Kimball model. We hope that Andrij enjoys seeing some of his old work re-examined in a new light. We certainly enjoyed preparing this publication in his honor.

\section*{Acknowledgements}

We first acknowledge the many wonderful collaborations with Andrij Shvaika and other members of his group during the past two decades. None of this work would have been possible without his help. We also acknowledge Lorenzo Del Re for a critical reading of the manuscript and for pointing out reference~\cite{parcollet} to us. Finally, this work was supported by
the Department of Energy, Office of Basic Energy Sciences, Division of Materials Sciences and Engineering under Contract No.~DE--SC0019126.
J.~K.~F. was also supported by the McDevitt bequest at Georgetown.



%
%
\ukrainianpart

\title{Педагогічне виведення динамічних сприйнятливостей}
%
\author{Р.Д. Нессельрод,  Дж.К. Фрірікс}
\address{
 Фiзичний факультет, Джорджтаунський унiверситет, вул. 37 \& О NW, Вашинґтон, округ Колумбiя 20057, США
}

\makeukrtitle

\begin{abstract}
Двочастинкові динамічні сприйнятливості є важливими для широкого класу різних експериментів у фізиці конденсованих систем і не тільки. Однак, більшість підручників уникає опису того як виводити ці функції відгуку, можливо через те, що вони розглядаються як занадто складні. В літературі більшість виведень стосуються узагальнених сприйнятливостей які є більш загальними, але і вимагають навіть ще вищого рівня складності. В цій роботі ми наводимо більш пряме виведення в контексті модельних гамільтоніанів, які можуть бути прямо спроектовані на модель домішки. Ми також представляємо альтернативне виведення для незвідної вершини в контексті моделі Фалікова-Кімбала.  
\keywords динамічна сприйнятливість, формалізм Келдиша, лінійний відгук
\end{abstract}

\end{document}